\documentclass{cjaa}                   % preprint, the final version for publication
\usepackage{graphicx}                   %for PS/EPS graphics inclusion, new
\usepackage{amssymb}                    %for PS/EPS graphics inclusion, new

\begin{document}

   \title{Detecting high-energy neutrinos from microquasars with
the ANTARES telescope}

   \volnopage{Vol.0 (200x) No.0, 000--000}      %%preserved for Editor. DOn't remove!
   \setcounter{page}{1}          %%starting page, preserved for Editor. DOn't remove!

   \author{G. F. Burgio\inst{1}\mailto{} for the ANTARES Collaboration\inst{2} }
   \offprints{G. F. Burgio}                   %% is disabled in fact

   \institute{INFN Sezione di Catania, Via S. Sofia 64, I-95123 Catania, Italy\\
             \email{fiorella.burgio@ct.infn.it}
       \and
             http://antares.in2p3.fr\\ }

   \date{Received~~2004 month day; accepted~~2004~~month day}

   \abstract{
The ANTARES project aims at the construction of a neutrino telescope 2500 m 
below the surface of the Mediterranean sea, close to the southern French coast.
The apparatus will consist of a 3D array of photomultiplier tubes,
which detects the Cherenkov light emitted by upward going neutrino-induced 
muons. High-energy neutrinos may be produced in powerful cosmic accelerators,
such as, gamma-ray bursters, active galactic nuclei,  
supernova remnants, and microquasars.
We have estimated the event rate in ANTARES of neutrinos coming 
from these sources, and particularly for a microquasar model, and found 
that for some of these sources the detection rate can be up to 
several events per year.
   }

 \authorrunning{G. F. Burgio for the ANTARES Collaboration}   %author_head in even pages
 \titlerunning{Detecting high-energy neutrinos from microquasars......}  % title_head in odd pages

   \maketitle
%% The author head (on even pages) and the title head (on odd pages) will be
%% automatically extracted from \author{} and \title{}. Whenever the title is too long,
%% you will be asked to supply a shorter one by inserting either \authorrunning{} or
%% \titlerunning{} before \maketitle. Anyway, you can specify your own heads in advance.
%%
%%
%% Note: In the following text body of your manuscript, please note several differences from
%%       other major journals:
%% (1) \subsection{Please Capitalize the First Letter of Each Notional Word in Subsection Title}
%% (2) Please Capitalize the First Letter of Each Notional Word in table's caption

%
%________________________________________________ sections below
%
\section{Introduction}           %% first-level sections will be auto-capitalized
\label{sect:intro}
%\hspace{15pt}%                   %% preserved for Editor
A promising challenge for exploring the Universe is the detection of 
high-energy ($\gtrsim 10~\rm GeV$) neutrinos. As a matter of fact, 
the early Universe
cannot be probed with high-energy photons due to photon-matter and 
photon-photon interactions - gamma rays of a few hundred TeV from the Galactic
Centre cannot survive their journey to the Earth.
The weakly interacting nature of neutrinos, and the fact that they are
not deflected by magnetic fields, make them unique 
``probes'' for exploring regions at distances larger than 50 Mpc.

Neutrinos may be produced by cosmic accelerators,
like those in supernova remnants, active galactic nuclei, microquasars, 
and gamma-ray bursts.
The commonly accepted model of neutrino production in sources
is the socalled ``{\it astrophysical beam dump}'' model, 
in which neutrinos and photons are produced in $pp$ or $p\gamma$ interactions 
of accelerated protons with matter or photons via pion decay. 
Neutrinos escape from the source and travel large distances to the Earth, 
thus delivering information directly from the sites of acceleration.

Another possible source of neutrinos is the annihilation of
neutralinos, expected to be gravitationally trapped at the centres
of massive bodies such as the Earth, the Sun or the Galaxy, thus
contributing to the dark matter quest.

Several projects are now underway to construct large-scale neutrino detectors 
underwater
or under ice, such as Baikal, AMANDA, NESTOR, ANTARES, NEMO, IceCube 
(Carr 2003, Halzen 2003), with Baikal and AMANDA 
already running. Those detectors are optimised for detecting muons from
charged-current reactions of muon neutrinos, but are also sensitive to 
other neutrino flavours and to neutral-current reactions.
Detectors are constructed at large depths where the atmospheric muon
flux is significantly reduced compared to that at the surface. Upward-going
muons, produced by neutrinos having crossed the Earth, are then recognised 
as the products of neutrino interactions in or close to the instrumented
region. 

The technique employed by the neutrino telescopes is dictated by the 
small neutrino cross section and the large background due to atmospheric 
muons. Natural Cherenkov radiators, such as water or ice, provide a large 
active volume at reasonable costs and the indirect detection of neutrinos 
through muons profits from the increase of the ``target'' region by the muon
 range.
The Cherenkov light emitted by charged particles in deep water or ice is 
detected using an array
of photomultiplier tubes (PMTs) which are housed, together with some associated
electronic components, in a pressure-resistant glass sphere known as an 
optical module (OM). The muon direction and energy are measured using 
the arrival times and amplitudes of the PMT pulses. The detector sensitivity 
increases with energy due to the increase in the hadron-neutrino cross 
section, the longer muon range and the 
increase in the amount of emitted Cherenkov light through secondary  particles.

%% ChJAA editors DID NOT use \cite{} for citation, \ref and \label for
%% cross-references of Table/Figure in publication version.
%% ChJAA editors prefered you giving a citation as 'Michel et al. 1992', and
%% writting Table~1 or Fig.~1 and so forth. However, that will make authors
%% inconvenient in adjusting/adding/removing text, tables or figures. Anyway,
%% authors can use \cite, \citep and \citet as widely used in other journals.
%% ChJAA editors are moving to use a more flexible LaTeX source.

\section{The ANTARES design}
\label{sect:Des}
%\hspace{15pt}%                   %% preserved for Editor

The ANTARES (Astronomy with a Neutrino Telescope and Abyss environmental
RESearch) project started in 1996 and involves physicists, astronomers,
sea science 
experts and engineers from France, Germany, Italy, Russia, Spain, The 
Netherlands and the United Kingdom. 
The ANTARES detector will be deployed at about 2500 m depth in the 
Mediterranean Sea, 37 km off-shore of La Seyne sur Mer, near Toulon (France). 
The ANTARES location ($\rm 42^o~50'~N, 6^o~10'~ E$) gives an annual 
sky coverage
of about $\rm 3.5\pi~sr$. ANTARES will look at the whole southern hemisphere, 
and a significant fraction of the northern hemisphere. The Galactic Centre,
an important potential source of high-energy neutrinos, will be visible for 
$\rm 67\%$ of the day. The instantaneous overlap with the AMANDA 
experiment, located at the South Pole (Ahrens et al., 2002), will be 
approximatively $\rm 0.5\pi~sr$,
and the total overlap will be $\rm 1.5\pi~sr$. Hence crosschecks will
be allowed over possible point sources.

An intense $\rm R\&D$ and Site Evaluation programme has provided the relevant 
environmental parameters of the detector site. Extensive surveys
of the water optical properties have been carried out, 
along with a detailed analysis of the optical background due to 
bioluminescence, biofouling on optical modules and light transmission 
properties.
 
The ANTARES telescope will detect the Cherenkov light emitted by secondary 
particles produced in neutrino interactions in sea water or rock below the sea
bed. The detector, illustrated in Fig. 1, consists of a 3-dimensional array
of OMs arranged in 12 lines made of mechanically resistant electro-optical 
cables. The lines will be 
anchored at the sea bed at distances of about 70 m one from each other, 
and tensioned by buoys at the top. 
Each line has a total length of about 450 m, of which about 350 m will be 
equipped with 75 
optical modules arranged in triplets (storeys, see Fig.1), and oriented 
at $\rm 120^o$ in azimuth apart from each other and at an
angle of $\rm 45^o$ below the equator, in order to be mostly sensitive to 
neutrinos which have crossed the Earth. The vertical distance between 
adjacent 
storeys will be 14.5 m. The whole apparatus will be connected to the 
onshore station by a 40 km long electro-optical cable (EOC), deployed in 
October 2001. The EOC terminates with a junction box, which transmits
power and sends the data to the shore station, where they are recorded.

%-------------------------------------------------------------
   \begin{figure}
   \centering
   \resizebox{0.9\textwidth}{!}                % ! -- default height
   {\includegraphics[]{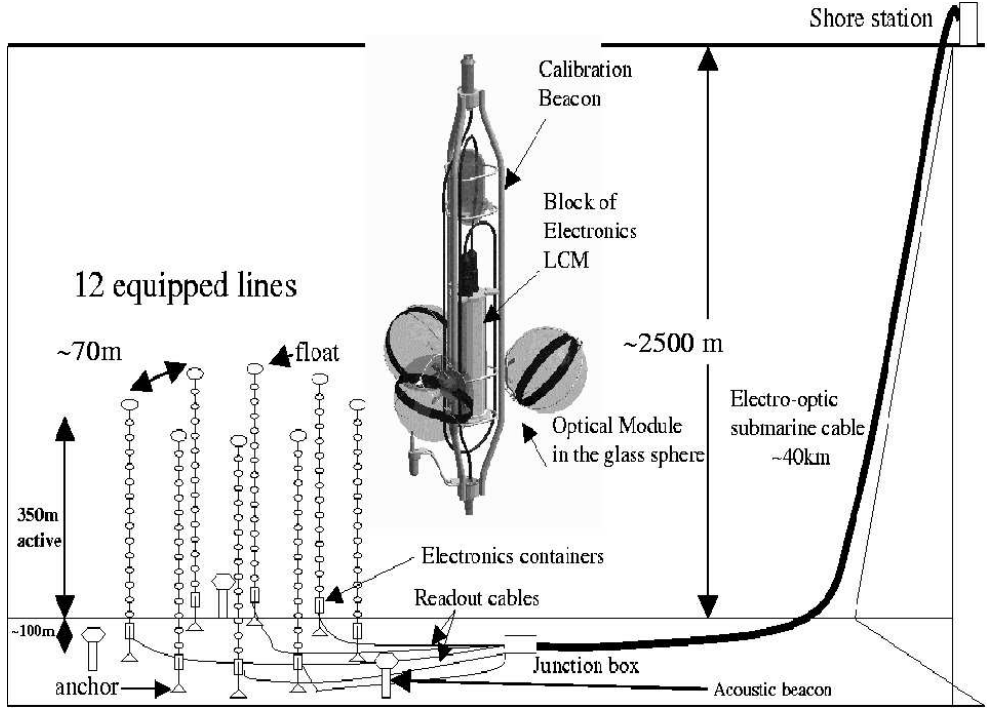}}    
   \caption{Schematic view of the ANTARES detector. }
   \label{Fig:layout}
   \end{figure}
%-------------------------------------------------------------

\section{Expected performance}
\label{sect:Perf}
%\hspace{15pt}%                   %% preserved for Editor

The ANTARES scientific programme is mainly devoted to the detection of 
neutrinos of astrophysical origin produced in point-like sources or coming 
from a distribution of sources in the whole sky which produces 
a diffuse flux. The neutrino signal can be detected 
above the background due to atmospheric neutrinos at energies greater
than 10-100 TeV, as a result of the harder neutrino spectrum expected from 
cosmic accelerators compared to neutrinos from cosmic ray interactions
in the atmosphere. 

The important parameters which characterize a neutrino telescope are the
effective area, the angular resolution and the energy resolution.
In particular, the effective area is crucial because it  
determines the event rate in a detector.
In fact, for point-like sources at a declination $\delta$ producing 
a differential flux of neutrinos $\rm \frac{d\Phi}{dE_{\nu}}$, 
the event rate is given by
\begin{equation}
N_{\mu}(\delta) = 
\int_{E_{min}}^{E_{max}} dE_{\nu} A^{eff}_{\nu}(E_{\nu}, \delta) 
\frac{d\Phi}{dE_{\nu}} \, ,
\label{eq:evrates}
\end{equation}
where $E_{min}$ and $E_{max}$ are the minimum and maximum energies
of neutrinos for the considered flux, respectively.
$A^{eff}_{\nu}$ is the sensitive area `seen' by $\nu$'s producing 
detectable $\mu$'s when entering the Earth.
The effective area depends on the neutrino energy, the efficiency of 
reconstruction and selection cuts.

\begin{figure}[h]
  \begin{minipage}[t]{0.5\linewidth}
  \centering
  \includegraphics[width=65mm, height=65mm]{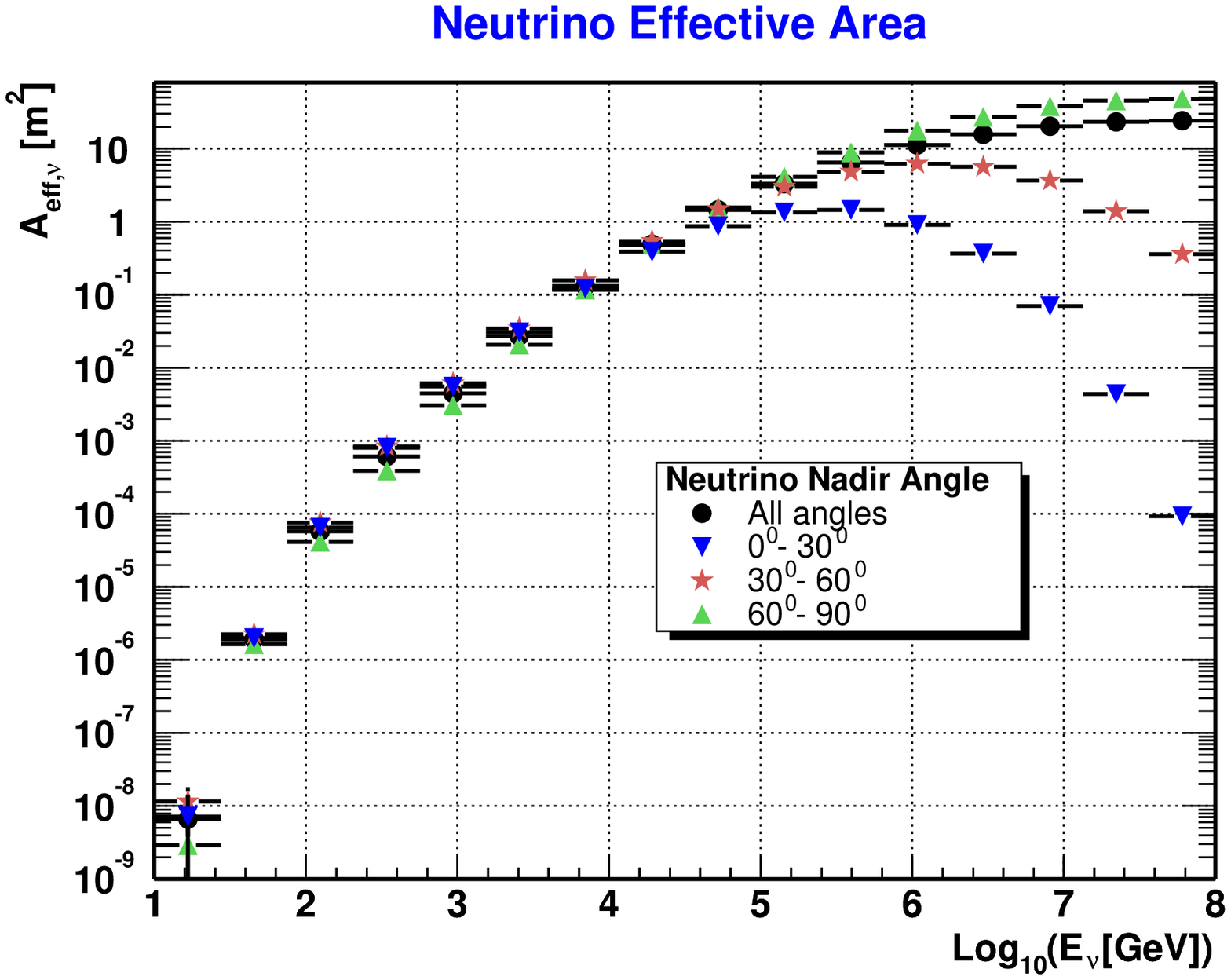}
  \vspace{-5mm}
  \caption{{\small Effective area vs neutrino energy after quality cuts:
circles are for a uniform event distribution, whereas the other curves
are for specific neutrino arrival directions.} }
  \end{minipage}
  \begin{minipage}[t]{0.5\textwidth}
  \centering
  \includegraphics[width=65mm, height=65mm]{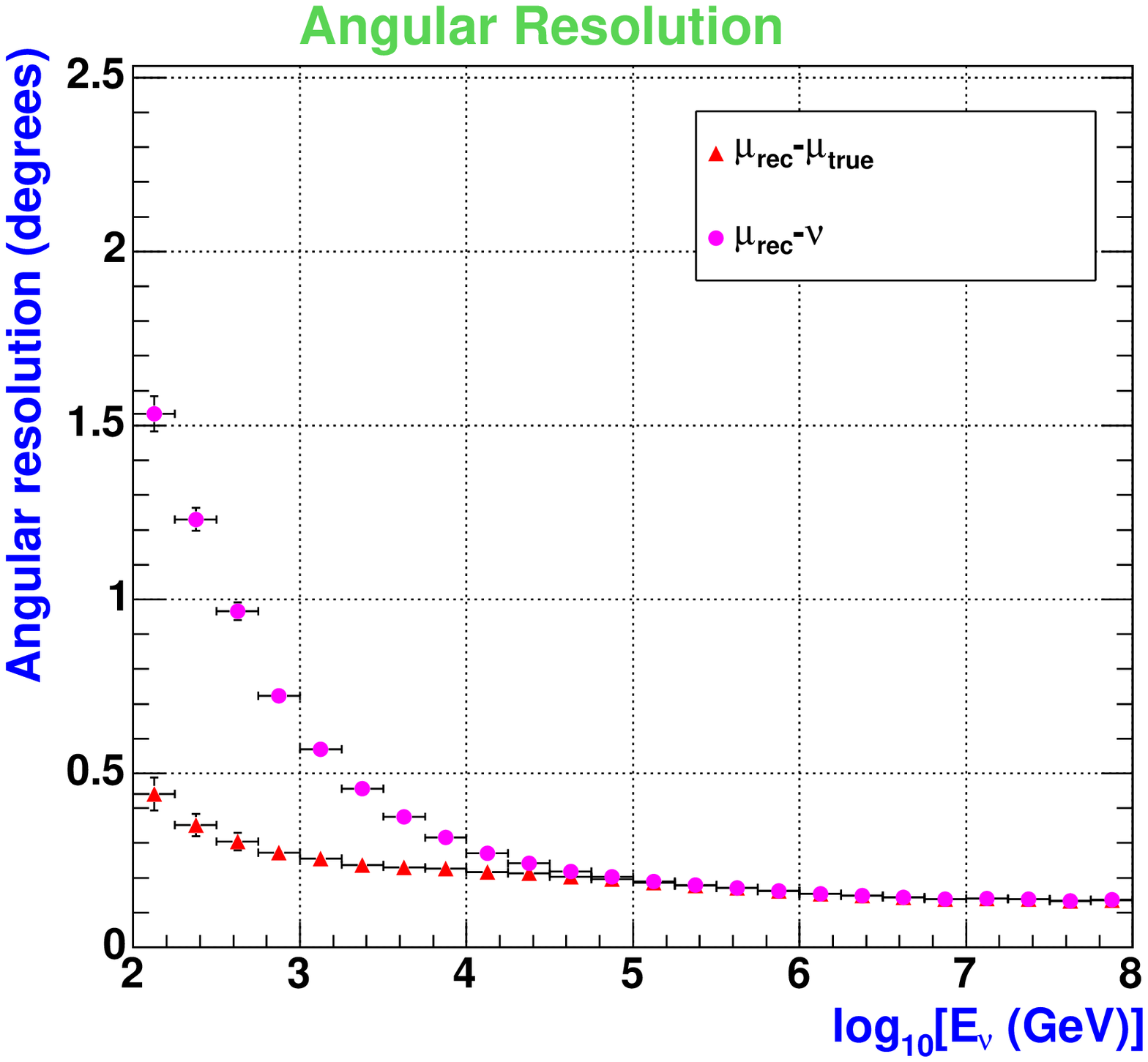}
  \vspace{-5mm}
  \caption{{\small Median angle of the distributions of the angle betwen
the reconstructed muon and the simulated muon (triangles) or the
simulated neutrino (circles) versus neutrino energy.}}
  \end{minipage}
  \label{Fig:fig23}
\end{figure}
In Fig. 2 we display the neutrino effective area as a function of the
neutrino energy. 
The figure shows the effective area for a uniform
event distribution (circles) and for three different angular ranges of 
the arrival directions. In particular, the downward triangles refer 
to arrival directions closer to the vertical one, 
whereas the upward triangles refer to directions closer to the horizon. 
The dependence on the arrival direction is due to the Earth's opacity.
The effective area increases rapidly up to about 1 PeV. Above this energy,
the effective area for a uniform event distribution saturates because 
of the balance between the increase of the hadron-neutrino cross section 
and the Earth's absorption. The latter strongly decreases the effective area 
of the detector for events around the vertical.

The intrinsic angular resolution of the telescope, defined as the median 
angular separation between the true and the reconstructed muon track, has been
estimated from simulation. In Fig. 3 the median value of the distribution of 
the angle
between the reconstructed and simulated muon is shown by triangles,
while the circles represent the median angle between the reconstructed 
muon and the parent neutrino as a function of the neutrino energy.
At smaller energies the angle between the 
reconstructed muon and the parent neutrino direction is dominated
by the kinematics of the charged current interaction. 
The improvement of the angular resolution with increasing energy results
in an increase of the signal to noise ratio for neutrino astrophysics 
studies at energies greater than $\approx$ 10 TeV. 
At these energies, the pointing
accuracy is only limited by the intrinsic angular resolution with a limiting 
value of $\rm 0.15^o$. 

The energy resolution in ANTARES and the methods of reconstructing muon energy
and parent neutrino spectra are discussed in Romeyer et al. (2003).
The muon energy, $E_\mu$, is determined from the muon range at small energies
and from the Cherenkov intensity due to radiative energy losses at
high energies. The resulting energy resolution is about a factor $\sim 2$
above 1 TeV. 

\section{Sensitivity to microquasars}

Microquasars have recently been considered as galactic candidates
for emission of high energy neutrinos. The observation of 
relativistic radio jets from some sources (Mirabel, 2004) 
has stimulated the idea that processes of neutrino production by relativistic
particles, similar to those believed to occur in active galactic nuclei 
(AGNs), can be considered.

Recently, Levinson \& Waxman (2001) have proposed a model for neutrino 
production in microquasars, similar to the AGN 
proton initiated cascade model of Mannheim et al.~(1992). 
The authors argue that hadrons can be accelerated up to $\sim 10^{16}$
eV in the inner part of the jet by internal shocks.
Pions are produced in collisions of hadrons with external X-ray photons,
and with the synchrotron photons from the jet produced by leptons 
accelerated in this same shock. The neutrino fluxes, expected in the energy 
range 1-100 TeV, are very high, especially from persistent sources, e.g.,
$\sim 10^3$ neutrino event rates in a 1 km$^2$ neutrino detector in the case 
of SS433. In a subsequent paper, Distefano et al. (2002) estimated the 
expected neutrino event rates in a 1 km$^2$ detector from the large
population of known microquasars,
assuming that relativistic protons take 10$\%$ of the jet power. 
The authors conclude that some microquasars with persistent activity, 
particularly SS433 and GX339-4, should produce more than 100 neutrino 
events per year in such a 1 km2 detector, while microquasars with 
episodic activity, e.g. Cyg X-3 or GRO J1655-40, might yield a few 
neutrino events per long-duration flare.  Therefore smaller detectors such 
as AMANDA-II or ANTARES might observe a small number of neutrino events 
from microquasars. 
%-------------------------------------------------------------
   \begin{figure}
   \centering
   \resizebox{0.9\textwidth}{!}                % ! -- default height
   {\includegraphics[]{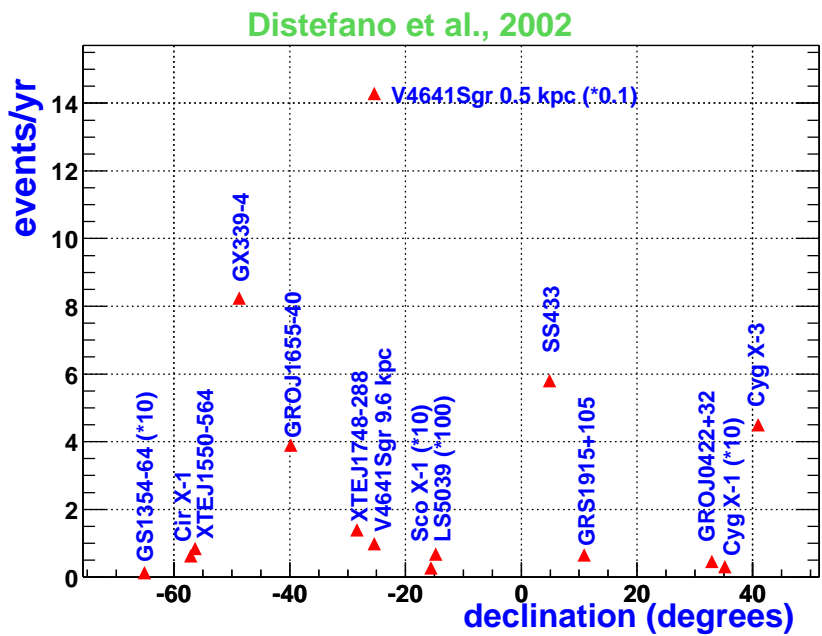}}    
   \caption{Event rate from known microquasars in ANTARES. }
   \label{Fig:mq}
   \end{figure}
%-------------------------------------------------------------
We have calculated the predicted number of events in 
ANTARES for some selected sources by convolving the neutrino fluxes 
given in Distefano et al. (2002) with the ANTARES effective area.  
In Fig. 4 we show the expected event rate in one year in ANTARES against the 
source declination. 
Four or five of the microquasars considered can produce a 
significant event rate, given an atmospheric neutrino background of the order 
of 0.5 events per year in $\rm 1^o$ around the source.
About 6-8  events per year are expected for the two persistent
sources, SS433 and GX339-4. In particular, GX339-4 will be  
always visible for ANTARES and is never seen by AMANDA, whereas SS433 is seen 
by both detectors (11.5 hr/day for ANTARES and 24 hr/day for AMANDA) and 
therefore provides a cross-check.

As clearly shown in Fig. 4, most of the sources will produce in ANTARES an 
event rate per year not larger than one.  For the source V4641 Sgr, 
two sets of parameters have been adopted: small distance (0.5 kpc)
and large angle of sight (63$^o$), which corresponds to the one shown 
in the figure, or larger distance (9.6 kpc) 
with a jet directed toward our line of sight (9$^o$). The latter leads
to a much higher neutrino detection rate, about $\sim 100$ events/year.

\section{Summary and Conclusions}

The ANTARES neutrino telescope is foreseen to be fully deployed by 2007.
The project has now entered the construction and deployment phase
for a 0.1 $\rm km^2$ scale detector.
The assembly and deployment of the first full line will be performed 
at the beginning of 2005. 
The $\rm R\&D$ phase of the project, which started in 1996, is now 
completed. During this phase a detailed assessment of the main requirements
for an undersea neutrino telescope was made. 
After the installation of the electro-optical cable and the
junction box, a prototype string and a string dedicated to environmental 
parameter measurements have been deployed, operated and recovered.
A large bulk of data was acquired and analysed, helping to identify and 
solve some problems. 

As far as the detection capability of ANTARES is concerned,
it has been found that the highest detection rates are obtained
for galactic sources, especially young supernova remnants and microquasars. 
For the latter ones, we have taken a theoretical
model, which has been recently published by Distefano et al. (2002), 
and calculated the event rates in ANTARES for 14 visible sources.
We have found that only two microquasars, i.e. SS433 and GX339-4,
can produce in ANTARES an appreciable detection rate in one year, 
about 6-8 events. Moreover, SS433 is visible both by ANTARES and AMANDA 
telescope, and therefore for this source a crosscheck is possible.
Other microquasars, taken into account by Distefano et al. (2002),
do not produce in ANTARES an event rate per year larger than one.

\label{lastpage}


\begin{thebibliography}{99}
%% you can type \apj for ApJ, \aap for A&A, \apss for Ap&SS, etc. Please consult
%% the macro cjaa.cls. You can also find them in aasguide.tex (AASTeX for ApJ, AJ, PASP)
%% Please follow the format of ChJAA's reference list
\bibitem[2002]{am02}
AMANDA Collaboration (D. Ahrens et al.), 2002, \prd, 66, 012005.

\bibitem[2002]{an02}
ANTARES Collaboration (P. Amram et al.), 2002, Nucl. Instrum. and Meth. in 
Phys. Res., 484, 369.

\bibitem[2003]{car03}
Carr J., 2003, Nucl. Phy. B (proc. suppl.), 118, 383.

\bibitem[2002]{Dist02}
Distefano C., Guetta D. , Waxman E.,  and Levinson A., 2002, \apj, 575, 378.

\bibitem[2003]{hal03}
Halzen F., 2003, Proc. 10th Intern. Workshop on Neutrino Telescopes
(Venice, Italy, 11-14 March 2003), 2, 345.

\bibitem[2001]{lw01}
Levinson A., Waxman E., 2001, \prl, 87, 171101.

\bibitem[1992]{MB92}
Mannheim K., Biermann P. L., 1992, \aap, 253, 21. 

\bibitem[2004]{mi04}
Mirabel F., see contribution to these proceedings.

\bibitem[2003]{rom03}
Romeyer A., Bruijn R., and Zornoza J., Proc. 28th ICRC, 
(Tsukuba, Japan), 2003, 1329. 

\end{thebibliography}
\end{document}